\documentclass[12pt,a4]{article}

\usepackage{amsmath}
\usepackage{graphicx}

\newcommand{\new}[1]{#1}

\DeclareMathOperator{\argmin}{argmin}

\newcommand{\myvec}[1]{{\ensuremath\underline{#1}}}

\newcommand{\pd}[2]{\frac{\partial #1}{\partial  #2}}

\usepackage{makeidx}
\usepackage{lineno}

\begin{document}

\title{Practical Introduction to Clustering Data\label{sec:clustering}}
\author{Alexander K. Hartmann, Institute of Physics,\\
University of Oldenburg, Germany}

\maketitle

\begin{abstract}Data clustering is an approach to seek for structure in
sets of complex data, i.e., sets of ``objects''. 
The main objective is to identify groups of objects which are similar to each
other, e.g., for classification. Here, an introduction
to clustering is given and three basic approaches are introduced:
the \emph{k-means}
algorithm, \emph{neighbour-based clustering}, and an \emph{agglomerative
clustering} method. For all cases, C source code examples are given,
allowing for an easy implementation.

This introduction originates (with a couple of modifications) from
section 8.5.6 of the book: A.K. Hartmann, \emph{Big Practical Guide
to Computer Simulations}, World-Scientifc, Singapore (2015).
\end{abstract}

\vspace*{5mm}

\section{Introduction}

\index{clustering|(}
\index{data clustering|(}

Often one wants to find similarities in a set 
of $n$ objects, characterized
by ``feature vectors''. We assume that the data
is given  by $n$ $d$-dimensional real-valued 
data points $\{\myvec {x}^{(i)}\}$ ($i=0,\ldots, n-1$), 
with each data point 
$\myvec {x}^{(i)} = ( x_1^{(i)},\ldots, 
 x_d^{(i)})^T$. Furthermore, let the data exhibit some substructure,
i.e., one can organize the data into groups, called \emph{clusters},
 such that the objects within
the groups are more similar to each other compared to objects belonging 
to different groups. Note that this is not a precise definition. In fact,
a good definition does not exisit. Thus, what is a good clustering always
depends on the application and on the data. This is already illustrated by
the two sample data sets A and B, which are shown in Fig.\
\ref{fig:cluster:sample}. For a detailed discussion of clustering see, e.g.,
Ref.\ \cite{jain1988}. Here, we discuss three approaches, the \emph{k-means}
algorithm, \emph{neighbor-based clustering}, and an \emph{agglomerative
clusering} method.

All C source code examples can be found in the file \texttt{cluster.c}
(main programm with simple examples plus subroutines). To compile
the program also the files \texttt{graphs.c}, \texttt{list.c},
and \texttt{graphs\_comps.c} plus corresponding header files must
be present. These files implement simple data structures and
algorithms for representing graphs, based on neighbour lists.
All files are provided in the with this \texttt{arXiv}
submission in the \texttt{anc} directory. Note that also the
\emph{GNU Scientific library has to be installed.} \cite{gsl}.
 A sample compile command is

\begin{verbatim}
cc -o cluster cluster.c graphs.c  list.c graph_comps.c \
   -lgsl -lgslcblas -lm -Wall
\end{verbatim}

\begin{sloppypar}
For using the main program, please have a look at \texttt{main()}
function in the \texttt{cluster.c} file. Note that  in \texttt{cluster.c} also
two functions \texttt{cluster\_test\_data1()} and 
\texttt{cluster\_test\_data2()} 
are included, which
generate the sample data sets used throughout this introduction.
\end{sloppypar}

\section{$k$-means Clustering}

\begin{figure}[!ht]
\begin{center}
\includegraphics[width=0.45\textwidth]{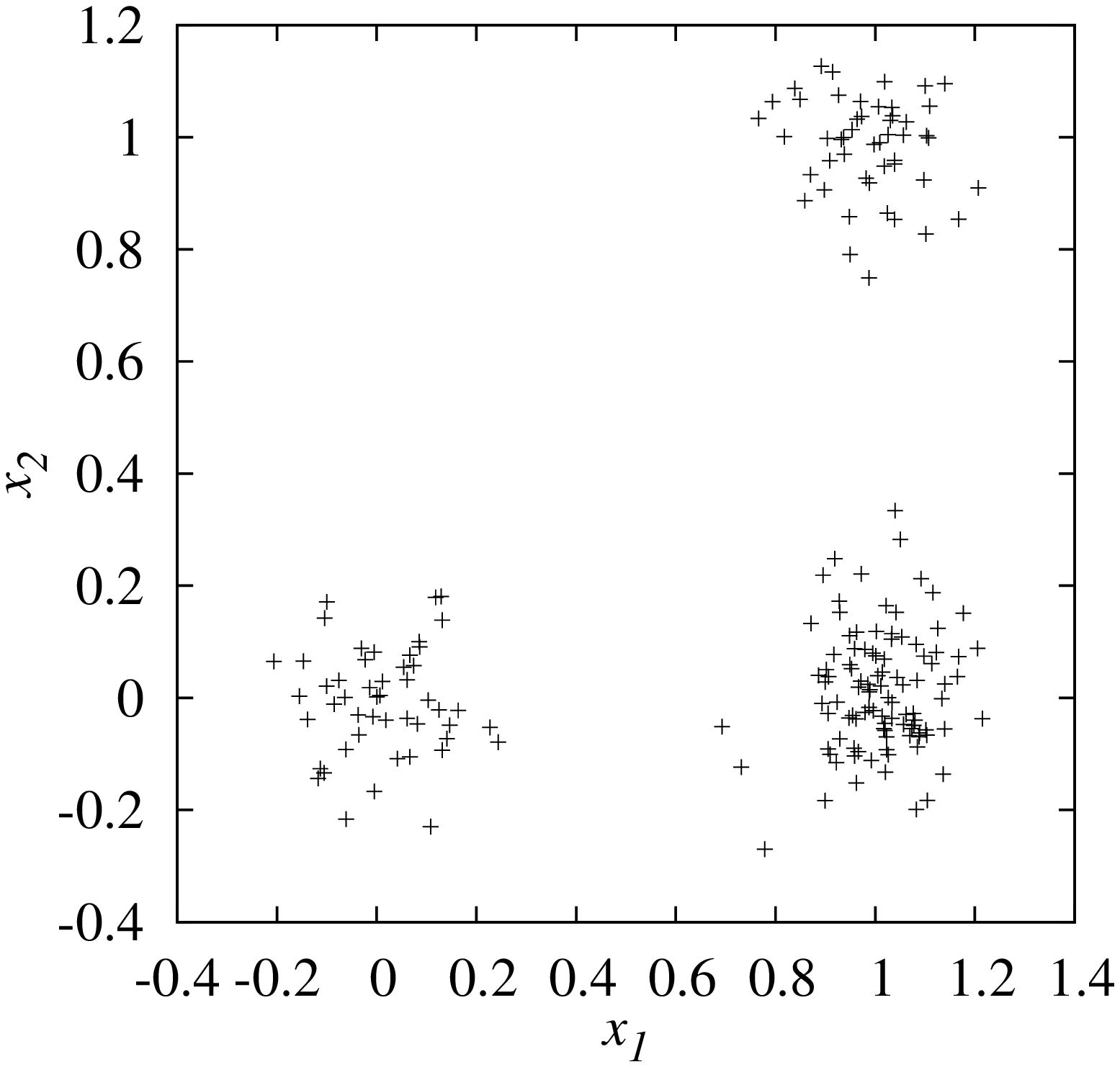}
\includegraphics[width=0.45\textwidth]{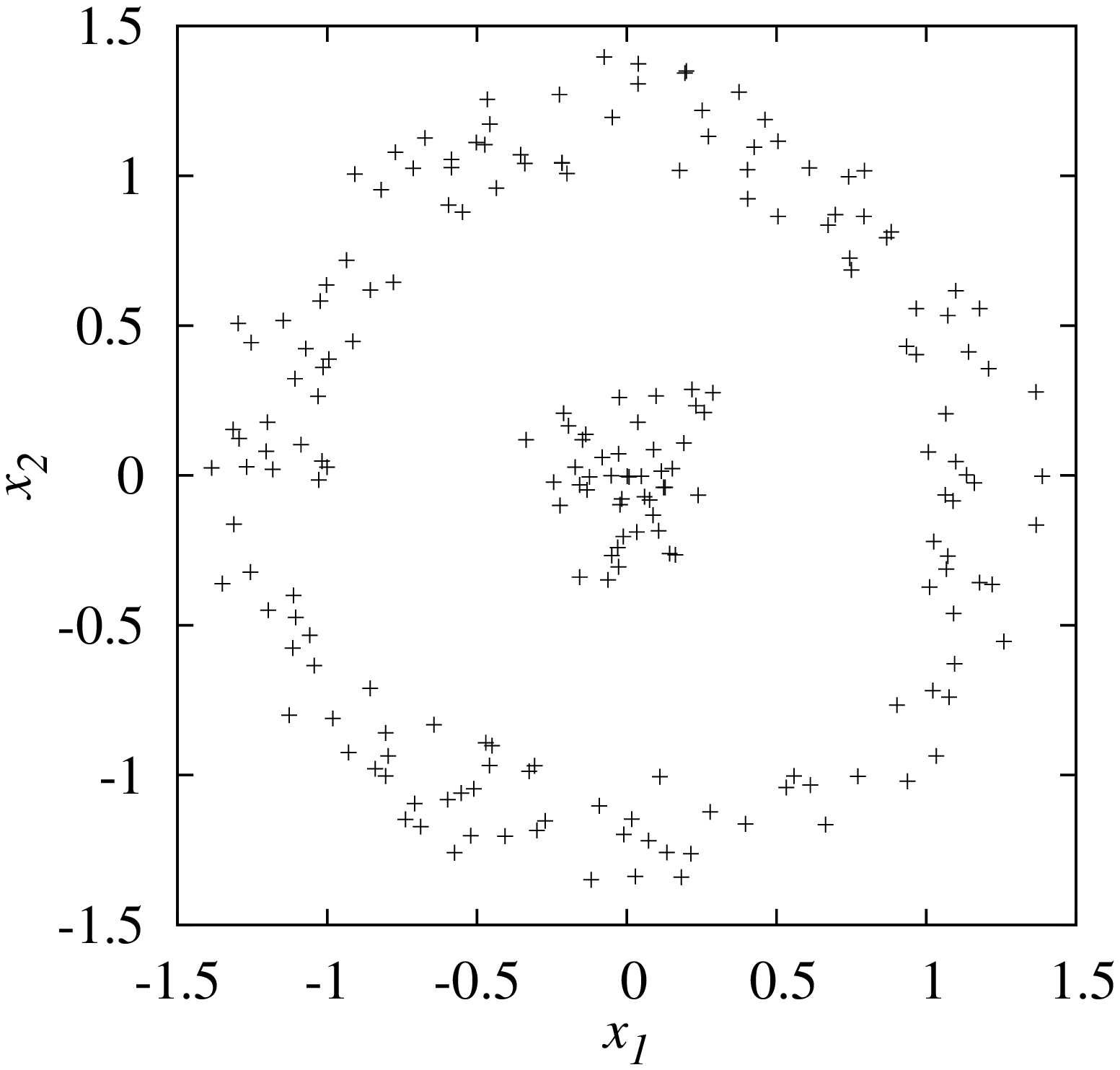}
\end{center}
\caption{Two sample sets A ($n=200$, left) and B ($n=200$, right) 
for sets of two-dimensional
data points, which will subsequently be used to test clustering algorithms,
aiming at identifying subsets of similar data points.
\label{fig:cluster:sample}}
\end{figure}

\index{clustering!k-means|(}

Following the \emph{$k$-means} approach, one wants to partition the data
set into $k$ clusters, $k$ being a somehow given parameter. Below we discuss
shortly the influence of the choice of $k$ on the resulting clustering. 
The $k$-means approach is based on a  geometric point of view. 
Each cluster $c=0\ldots,k-1$ 
shall be represented by a \emph{center}
vector $\myvec{\xi}^{(c)}$. 
Let us assume that each data point with index $i\in\{0,\ldots, n-1\}$
is assigned to some  (initially possibly randomly chosen) 
cluster $c(i)\in\{0,\ldots, k-1\}$.

We calculate the mean-squared difference (or ``spread'') $\chi^2$ of all
data points to the center of its cluster:
\begin{equation*}
\chi^2 = \frac 1 n \sum_{i=0}^{n-1} 
\left(\myvec{\xi}^{(c(i))}-\myvec{x}^{(i)}\right)^2\,.
\end{equation*}
We assume that the best choice of the center vectors and of the
assignment to the clusters is the one which minimizes the spread.
Thus, for a fixed assignment of data points to clusters and any cluster
$c\in\{0,\ldots,k-1\}$ we have for each direction $a\in\{1,\ldots,d\}$
the condition that the partial derivative of the spread with respect to the
$a$'th component of the center vector $\myvec{\xi}^{(c)}$ vanishes: 
\begin{equation*}
0 \stackrel{!}{=}
\pd{\chi^2}{\xi_a^{c}}= \frac 2 n \sum_{i=0}^{n-1} \delta_{c,c(i)} 
\left(\xi^{(c(i))}_a-x^{(i)}_a\right)
= 2 \frac {n_c}{n} \xi^{(c)}_a - \frac 2 n \sum_{i=0}^{n-1} \delta_{c,c(i)} 
x^{(i)}_a\,,
\end{equation*}
where $n_c=\sum_{i=0}^{n-1} \delta_{c,c(i)}$ is the size of cluster $c$.
Thus, each center vector $\myvec{\xi}^{(c)}$ is, 
as the name suggest, the geometric center
of the data points assigned to cluster $c$:

\begin{equation}
\myvec{\xi}^{(c)} = \frac 1 {n_c} \sum_{i=0}^{n-1} \delta_{c,c(i)} 
\myvec{x}^{(i)}\,.
\label{eq:k:mean:centers}
\end{equation}

On the other hand, 
for fixed centers $\myvec{\xi}^{(c)}$, minimizing $\chi^2$
 can be achieved by assigning
each data point to its closest cluster:
\begin{equation}
c(i) = \argmin_{c=0,\ldots,k-1} \left\{
\left(\myvec{\xi}^{(c)}-\myvec{x}^{(i)}\right)^2\right\}\,.
\label{eq:k:mean:clusters}
\end{equation}
Thus, a very simple algorithm can be obtained by starting with a random 
assignments of the data points to clusters and then iterating Eqs.\
(\ref{eq:k:mean:centers}) and (\ref{eq:k:mean:clusters}) until convergence,
e.g., until the the relative change of the center vectors is 
less than a small given threshold 
$\epsilon$. Note that this approach does \emph{not} guarantee
a convergence to a solution where the spread $\chi^2$ assumes its
\emph{global} minimum. See below for an example.

\index{GNU scientific library|(}
\index{library!GNU scientific|(}

Next, we discuss a short C implementation of the $k$-means approach.
Note that the file {\tt cluster.c} also contains auxiliary and test
functions, like {\tt cluster\_test\_data1()} and
{\tt cluster\_test\_data2()} which generate the test sets A and B,
respectively.
You can just use the code it as it is, or use it as a starting point for
a more refined approach, e.g., by introducing additional weights signifying the
importance of the data points.
The function {\tt cluster\_k\_means()} receives a matrix {\tt data}, which
contains the data points as column vectors, and the number $k$
of clusters. For convenience,
we use the data types
{\tt gsl\_vector} and  {\tt gsl\_matrix} of the \emph{GNU Scientific
Library (GSL)} \cite{gsl}, see also Sec.\ 7.3. of 
Ref.\ \cite{bigPracticalGuide2015}. Also we use
a GSL random number generator {\tt rng} for the initial assignment of
the data points to the clusters. The function returns
an array, which contains for each data point an integer specifying its cluster.
The array is created inside the function. Furthermore, the function returns
the final spread, via a pointer {\tt spread\_p} which is passed as argument.

{\small 
\linenumbers[1]
\begin{verbatim}
int *cluster_k_means(gsl_matrix *data, int k, gsl_rng *rng, 
                     double *spread_p)
{
  int *cluster;             /* holds for each point its cluster ID */
  gsl_matrix *center;         /* holds for each cluster its center */
  int *cluster_size;         /* holds for each cluster its #points */
  int dim;           /* number of components of data point vectors */
  int num_points;                         /* number of data points */
  int t, d, c;                                    /* loop counters */
  double spread, spread_old;          /* total distance to centers */
  double dist, dist_min;   /* (minimum) dist. between point/center */
  double diff;            /* lateral distance between point/center */
  int c_min;                 /* center which is closest to a point */
  int do_print = 0;                               /* for debugging */
\end{verbatim}
\nolinenumbers} 

\noindent
\new{For initializing, the number {\tt num\_points} of data points 
and the number {\tt dim} of entries are take from the GSL matrix
data structure (lines 15 and 16).  
Using this, the array {\tt cluster}, which is returned,
the array {\tt cluster\_size}, which holds for each cluster the
number of assigned data points, and a GSL matrix for the centers are allocated
(lines 17--19).
Also, each data point is assigned initially 
to a randomly chosen cluster (lines 21,22):
}

{\small 
\linenumbers[15]
\begin{verbatim}
  num_points = data->size2;                          /* initialize */
  dim = data->size1;
  cluster = (int *) malloc(num_points*sizeof(int));
  cluster_size = (int *) malloc(k*sizeof(int));
  center = gsl_matrix_alloc(dim, k);

  for(t=0; t<num_points; t++)    /* intial assignments to clusters */
    cluster[t] = (int) k*gsl_rng_uniform(rng);
  spread = 1e100;
  spread_old = 2e100;
\end{verbatim}
\nolinenumbers} 

\noindent
\new{
The main loop (lines 25--64) 
is performed until the spread changes by less than one percent (line 25).
In each iteration, for the given assignments of the data points to clusters,
the cluster sizes and the cluster centers  
are updated (lines 27--42) according to Eq.\ (\ref{eq:k:mean:centers}).
 This is achieved by first initializing centers
and cluster sizes to zero (lines 27--29), by next iterating over all
data points (lines 30--37), and by finally normalizing the centers
by the cluster sizes $n_c$ (lines 38--42). Note that in C,
the entries $1,\ldots, d$ of the data points run from 0 to {\tt dim}$-1$.

For each iteration, second, for each data point its closest cluster is
determined and the spread is recalculated (lines 44--63). This
involves in particular iterating for each data point over all
cluster centers (lines 48--62), determining the distance between the data
point and a center (lines 50--55) and 
determining the closest center (lines 56--60).
}
\newpage

{\small 
\linenumbers[25]
\begin{verbatim}
  while ( (spread_old-spread)>0.01*spread_old)        /* main loop */
  {
    gsl_matrix_set_all(center, 0.0);
    for(c=0; c<k; c++)
      cluster_size[c] = 0;
    for(t=0; t<num_points; t++)               /* determine centers */
    {
      cluster_size[cluster[t]]++;
      for(d=0; d<dim; d++)
        gsl_matrix_set(center, d, cluster[t], 
                       gsl_matrix_get(center, d, cluster[t])+
                       gsl_matrix_get(data, d, t));
    }  
    for(c=0; c<k; c++)
      if(cluster_size[c] > 0)
      for(d=0; d<dim; d++)
        gsl_matrix_set(center, d, c, 
                       gsl_matrix_get(center, d, c)/cluster_size[c]);

    spread_old = spread;  spread = 0;
    for(t=0; t<num_points; t++)        /* determine closest center */
    {
      c_min = -1;
      for(c=0; c<k; c++)                  /* test with all centers */
      {
        dist = 0;               /* calculate distance point/center */
        for(d=0; d<dim; d++)
        {  
          diff = gsl_matrix_get(center,d,c)-gsl_matrix_get(data,d,t);
          dist += diff*diff;
        }
        if( (c_min == -1)||(dist_min > dist))  /* closest center ? */
        {
          c_min = c;
          dist_min = dist;
        }
      }
      cluster[t] = c_min;  spread += dist_min;
    }
  }
\end{verbatim}
\nolinenumbers} 

\noindent
\new{At the end of the function, 
the current spread is stored in the external variable which
is given by the pointer {\tt spread\_p}. Also the memory for the
center vectors and the cluster sizes is freed and finally the {\tt cluster}
array containing the result is returned:}

{\small 
\linenumbers[65]
\begin{verbatim}
  *spread_p = spread;
  gsl_matrix_free(center);
  free(cluster_size);
  return(cluster);
}
\end{verbatim}
\nolinenumbers} 

\index{GNU scientific library|)}
\index{library!GNU scientific|)}

\begin{figure}[!ht]
\begin{center}
\includegraphics[width=0.45\textwidth]{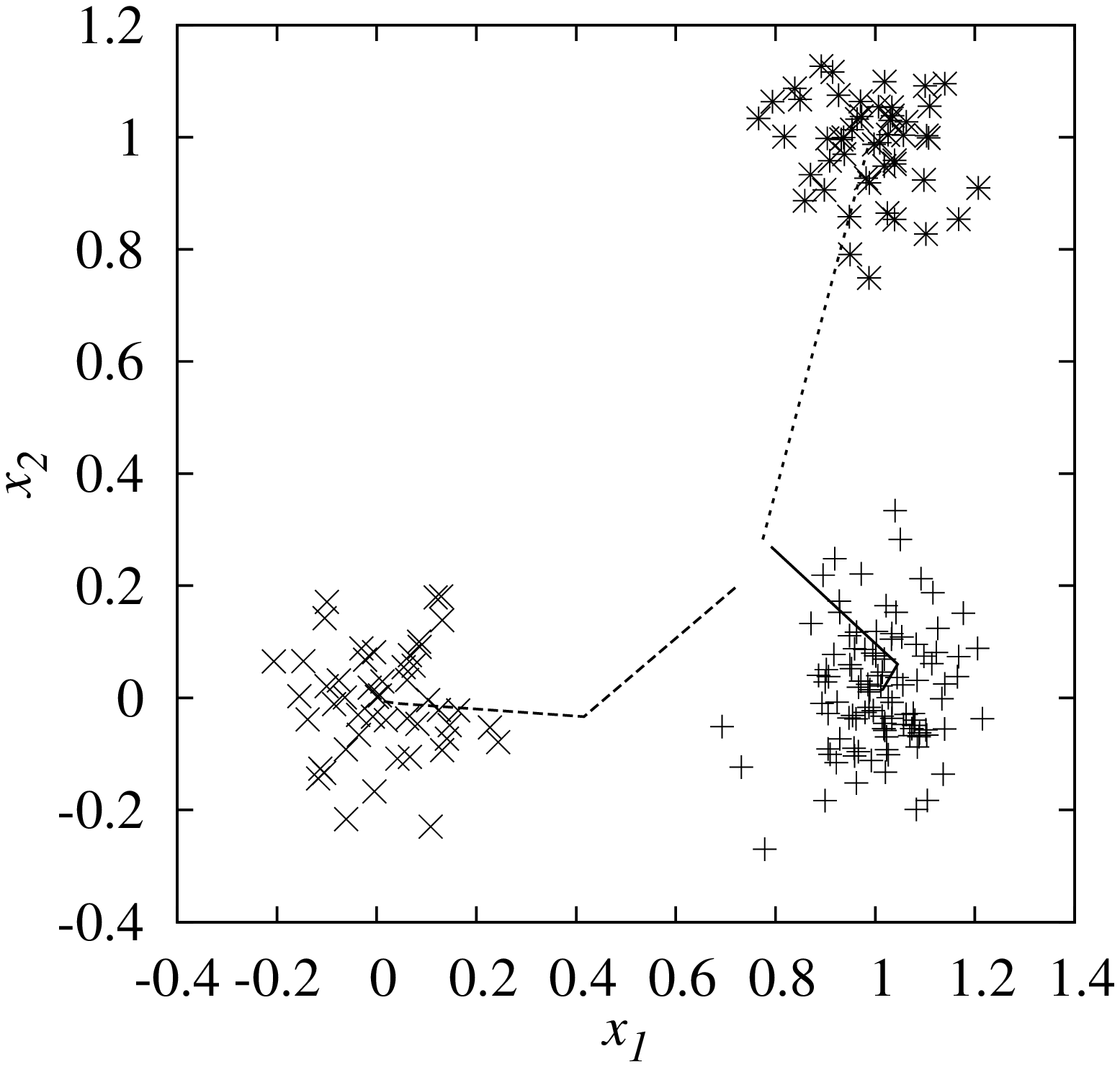}
\includegraphics[width=0.45\textwidth]{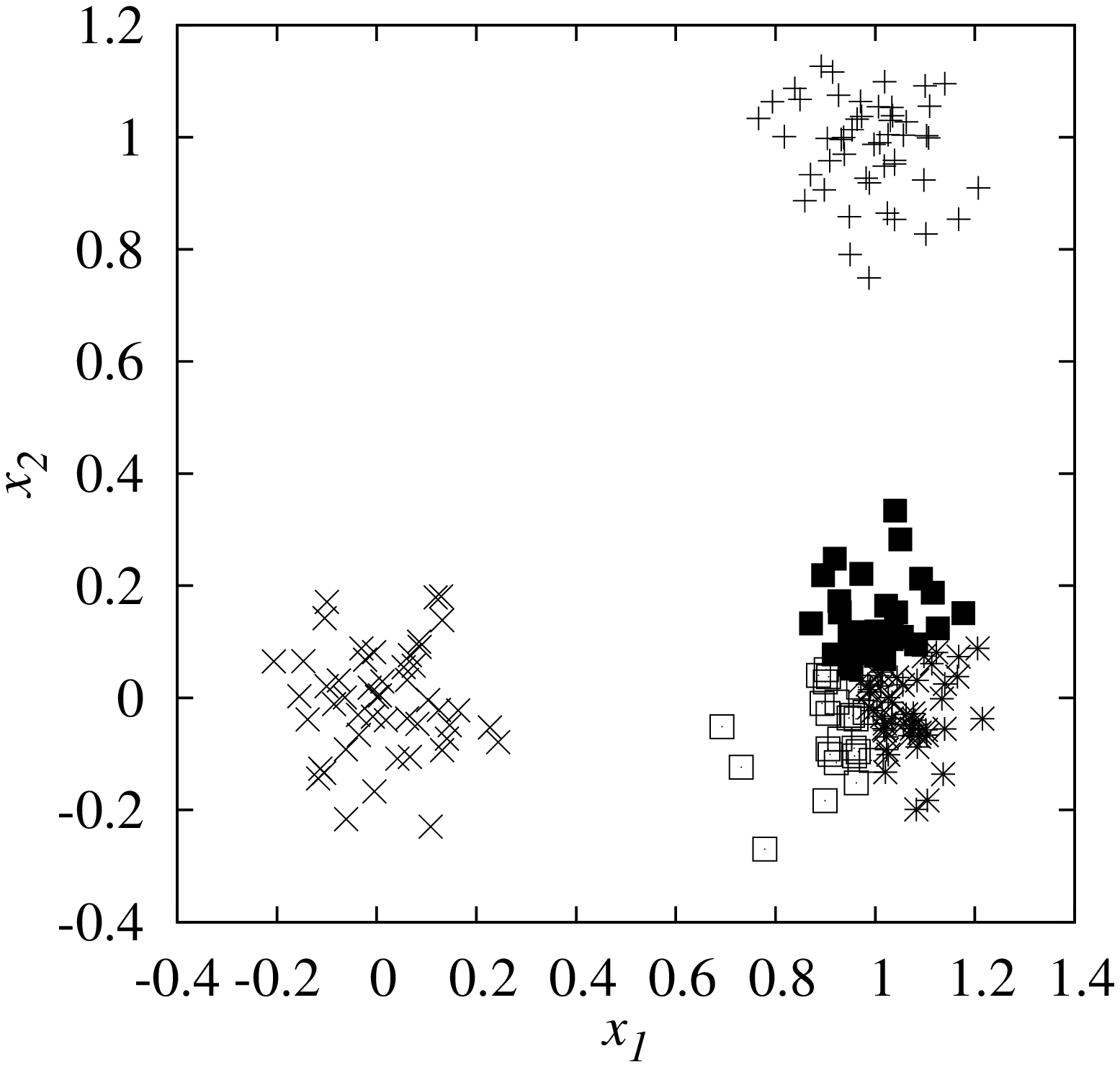}

\includegraphics[width=0.45\textwidth]{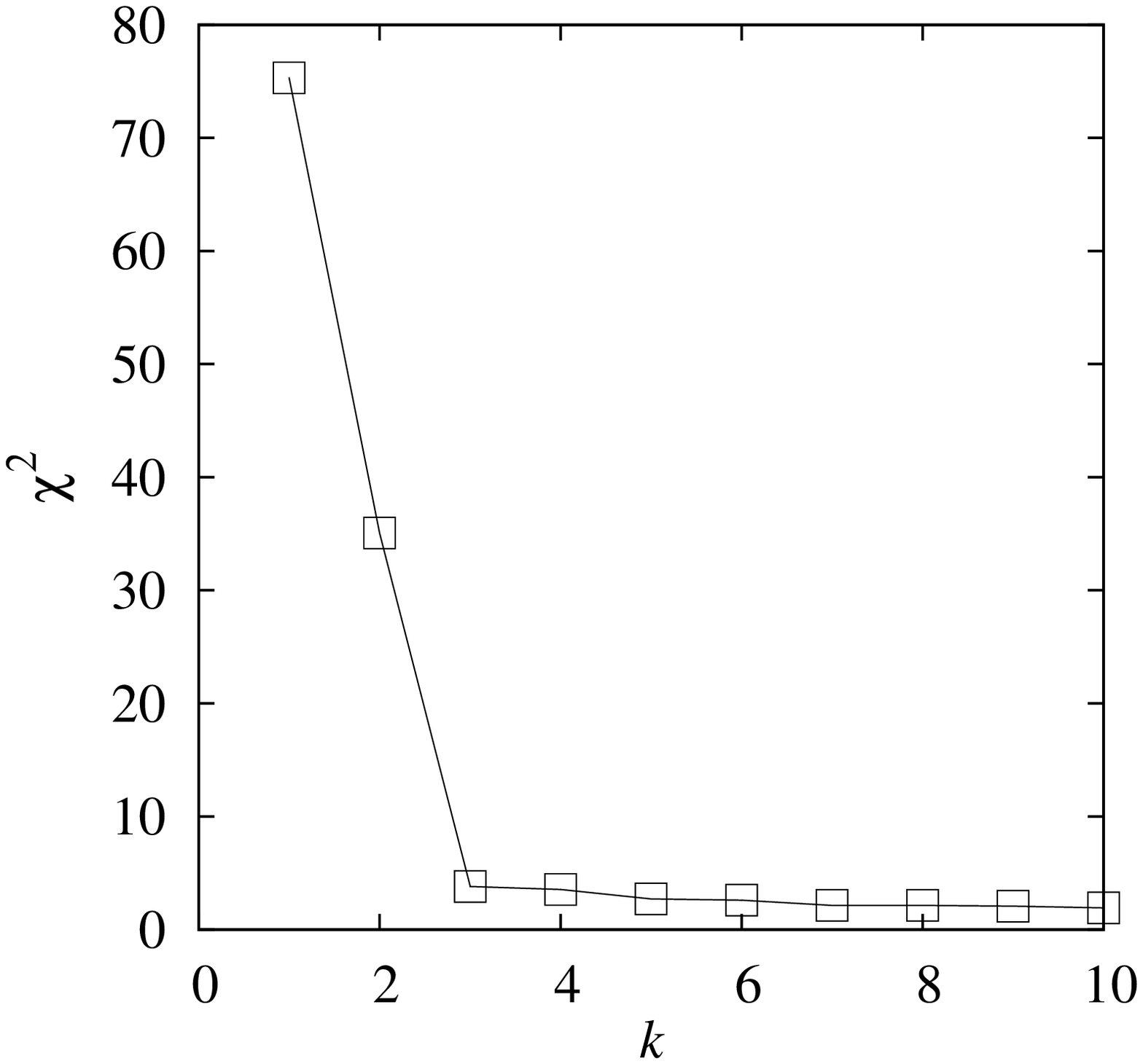}
\includegraphics[width=0.45\textwidth]{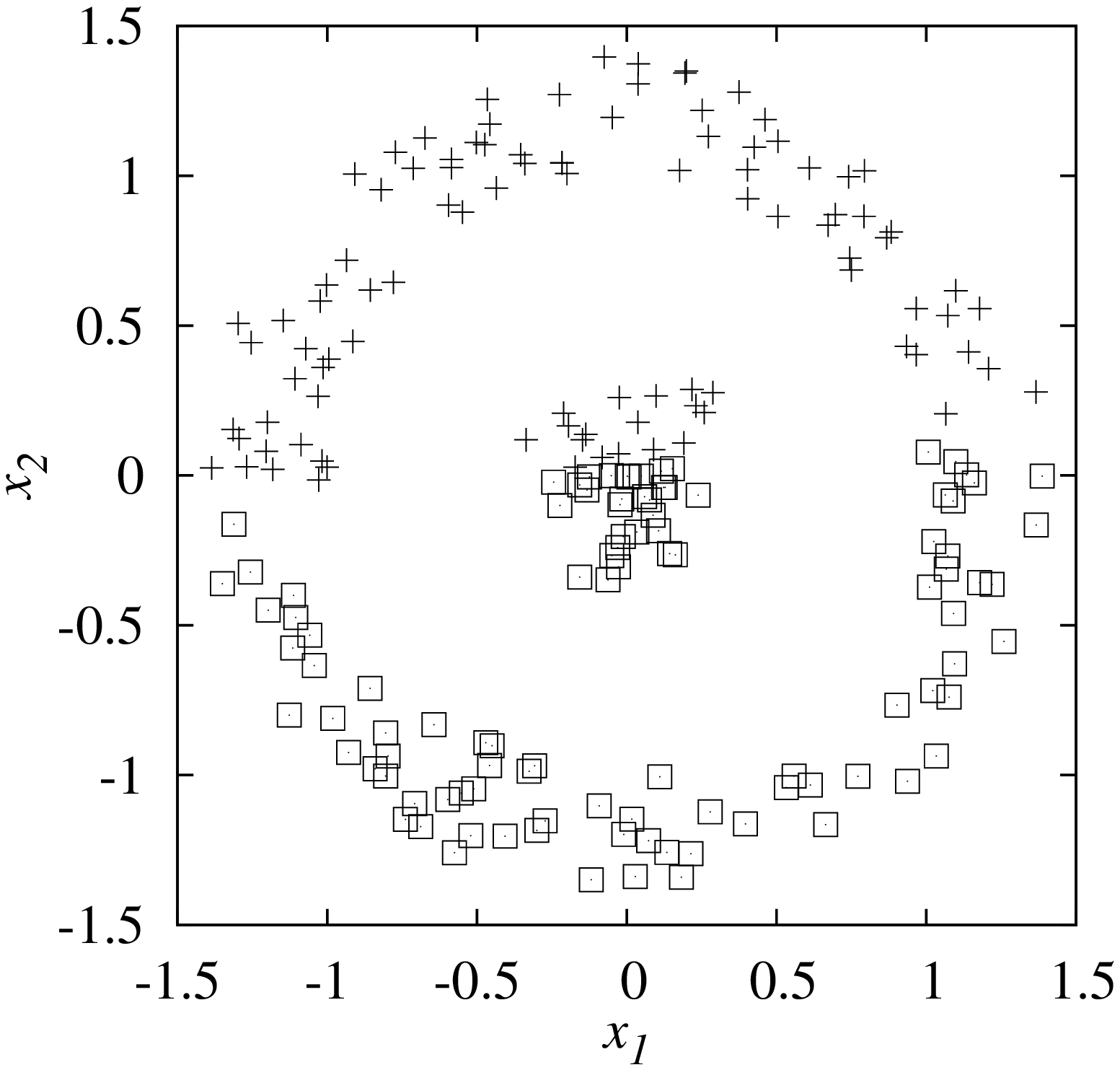}
\end{center}
\caption{
Upper left: Result of the clustering of sample set A with
the $k$-means algorithm for $k=3$. Different symbols correspond to different
clusters.
The lines show how the centers have moved during
the iterations of the algorithm. Upper right: result of the $k$-means
algorithm for sample set B and $k=5$. Here, the algorithm mistakenly subdivides
the cluster around $(1,0)^T$ into three sub clusters. Lower left:
spread $\chi^2$ as function of the number of clusters $k$. Above the
most suitable number $k=3$ the spread decreases only slightly when increasing
the number of clusters. Lower right: For sample set B, $k$-means fails
even if the most suitable number $k=2$ is chosen.
\label{fig:k:means}}
\end{figure}

\new{
In the upper left of Fig.\ \ref{fig:k:means} the result for the cluster
analysis of data set A is shown for the choice $k=3$. Also shown are
the ``paths'' the centers have taken during the iteration of the algorithm.
Obviously, the clustering represents the structure of the data well. This
changes in case the value of $k$ does not represent the data well, see
upper right of Fig.\ \ref{fig:k:means}, where the result for $k=5$ is shown.
Since the algorithm is forced to have five clusters, it subdivides
the cluster around $(1,0)^T$  into three clusters.
This case where $k$ is not well adapted serves also as an example 
to show that the simple iterative algorithm does not necessarily
converge to the global minimum spread. When repeating the clustering
for $k=5$ with different seeds for the random number generator,
different spreads and thus different cluster assignments will occur.
Such a non-unique convergence, observed after restarting the 
{\tt cluster\_k\_means()} function, may also be used as 
an indicator that $k$ is not well chosen.

\section{Neighbor-based clustering}

Often, the most suitable number $k$ of clusters is in fact not known in advance.
In this case, it helps sometimes to perform the clustering
for several values of $k$ and observe the spread $\chi^2$ as a function
of $k$, see lower left of Fig.\ \ref{fig:k:means}. The spread
shrinks monotonously when increasing $k$. When the spread
does not decrease significantly any more, a suitable number of clusters
is found. Nevertheless, this does not work always, e.g., when the clusters
exhibit a sub-cluster structure. 

There are also cluster structures, where the basic assumption that
each cluster can be represented by its geometric center fails, as
for sample set B (right of Fig.\ \ref{fig:cluster:sample}) where
two cluster are present. Whatever
value of $k$ is chosen, the $k$-means algorithm will not converge to the 
correct result. The reason is that both clusters, although being quite 
distinct, exhibit very similar geometric means. Here, clustering approaches
are needed, which take the local neighbor relations of data points into 
account, rather than the global positions of the data points.
}

\index{clustering!k-means|)}

\index{clustering!neighbor-based|(}

\new{As a first step, one needs for all pairs $i,j$ of data points
the notion of a ``distance'' $d(i,j)$. The best choice  for a distance
function depends heavily on the data set and the nature of the
clustering problem. For the sample sets A and B 
(see Fig.\ \ref{fig:cluster:sample}), which are just points
in the two-dimensional plane, the Euclidean distance appears
to be suitable:
\begin{equation}
d(i,j) \equiv \sqrt{\sum_{a=1}^{d} \left(x_a^{(i)}-x_a^{(j)}\right)^2}
\label{eq:distance:euclidean}
\end{equation}
Next, we present a C function which turns
the set of data vectors into a matrix of pair-wise distances.
The function receives a matrix {\tt data} 
(GSL data type {\tt gsl\_matrix})
of column vectors and returns a matrix of pair-wise distances.
Note that the number of data points and the dimensions, i.e.,
the number of entries, can be taken from the matrix {\tt data} (lines 9 and
10). The main loop over all pairs of data points is performed
in lines 13--24. The calculation of the distance is done in lines
16--21. As usually in C,
the elements $1,\ldots, d$ of the data points are stored in
entries 0 through {\tt dim}$-1$.
}

{\small 
\linenumbers[1]
\begin{verbatim}
gsl_matrix *cluster_distances(gsl_matrix *data)
{
  gsl_matrix *dist;                /* matrix containing distances  */
  int dim;           /* number of components of data point vectors */
  int num_points;                         /* number of data points */
  int t1, t2, d;                                  /* loop counters */
  double distance, diff;           /* auxiliary distance variables */

  dim = data->size1;
  num_points = data->size2;                          /* initialize */
  dist = gsl_matrix_alloc(num_points, num_points);

\end{verbatim}
\newpage
\begin{verbatim}
  for(t1=0; t1<num_points; t1++)         /* iterate over all pairs */
    for(t2=0; t2<=t1; t2++) 
    {
      distance = 0;
      for(d=0; d<dim; d++)                   /* calculate distance */
      {
        diff = gsl_matrix_get(data,d,t1) - gsl_matrix_get(data,d,t2);
        distance += diff*diff;
      }
      gsl_matrix_set(dist, t1, t2, sqrt(distance));         /* set */
      gsl_matrix_set(dist, t2, t1, gsl_matrix_get(dist, t1, t2));
    }

  return(dist);
}
\end{verbatim}
\nolinenumbers} 

\new{The basic idea of the \emph{neighbor-based clustering} is
to translate the data set into a graph \cite{GR-bollobas98,GR-swamy91}, this
is a set of objects (\emph{nodes}) and a set of pairs ob objects, 
i.e., connections (also called \emph{links} or
\emph{edges}). The translation of the data set into a graph works as follows:
For each data point $\myvec{x}^{(i)}$, there is a node $i$ in the graph.
 Furthermore, all pairs $i,j$ of nodes
are connected by an (undirected) edge $\{i,j\}$, 
if the distance between the corresponding
data points is smaller than some given threshold $\theta$, i.e.,
if $d(i,j) < \theta$. This is achieved by the following function,
which uses the graph data structures as found in the header file
\texttt{graphs.h}, which is included with this
text. The function receives the matrix of
distances and the  threshold value $\theta$. The code is
rather concise, because
one needs only to determine
the number of nodes (line 8), set up the nodes of the graph
(line 9) and iterate over all pairs of nodes to set an edge whenever
the distance is below the threshold (lines 10--13):
}
\newpage
{\small
\linenumbers[1]
\begin{verbatim}
gs_graph_t *cluster_threshold_graph(gsl_matrix *distance, 
                                    double threshold)
{
  gs_graph_t *g;
  int num_nodes;
  int n1, n2;                                      /* node counter */

  num_nodes = distance->size1;
  g = gs_create_graph(num_nodes);
  for(n1=0; n1<num_nodes; n1++)    /* loop over all pairs of nodes */
    for(n2=n1+1; n2<num_nodes; n2++)
      if(gsl_matrix_get(distance, n1, n2) < threshold)   /* edge ? */
        gs_insert_edge(g, n1, n2);                

  return(g);
}
\end{verbatim}
\nolinenumbers}

\begin{figure}[!ht]
\begin{center}
\includegraphics[width=0.45\textwidth]{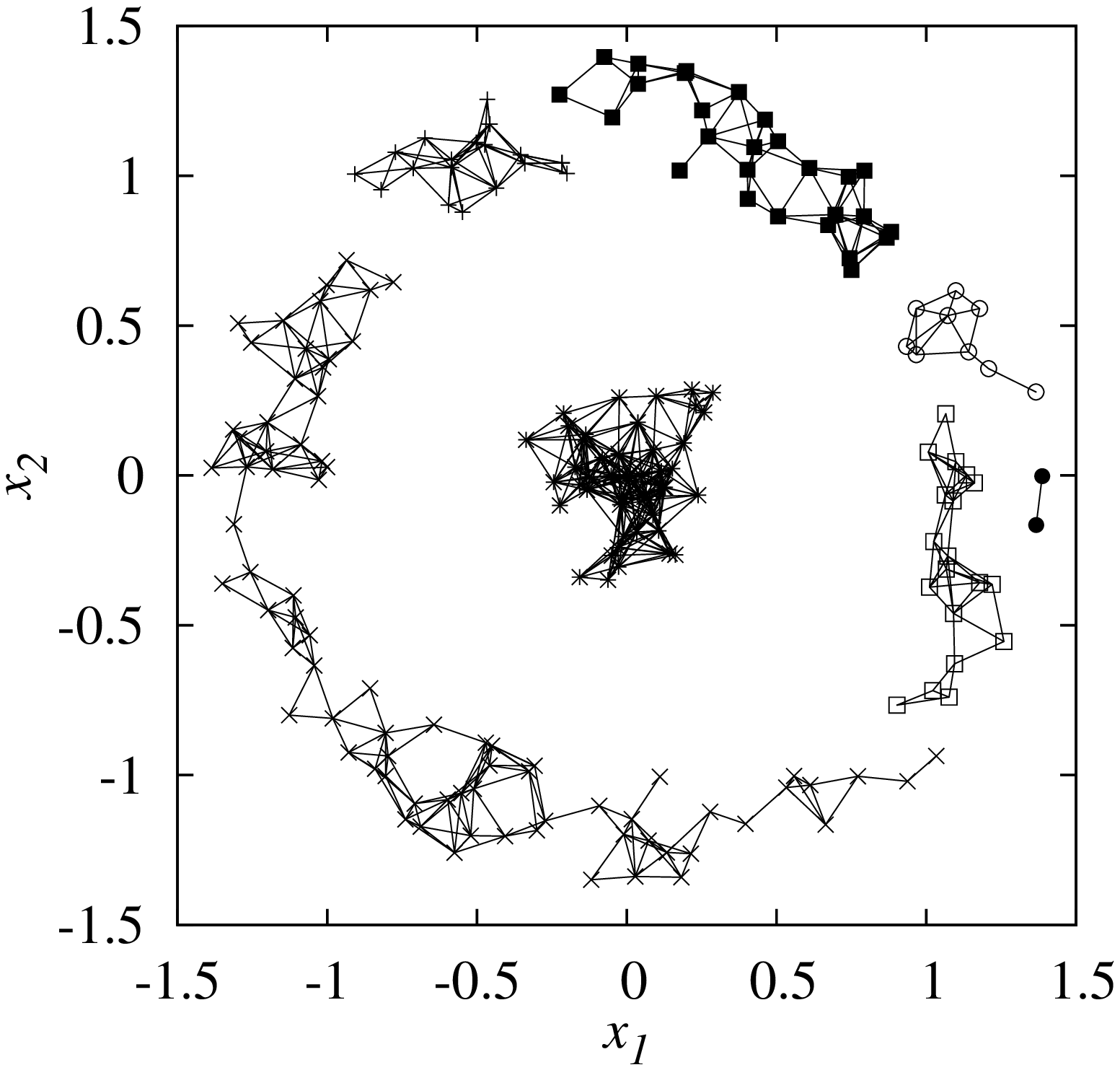}
\includegraphics[width=0.45\textwidth]{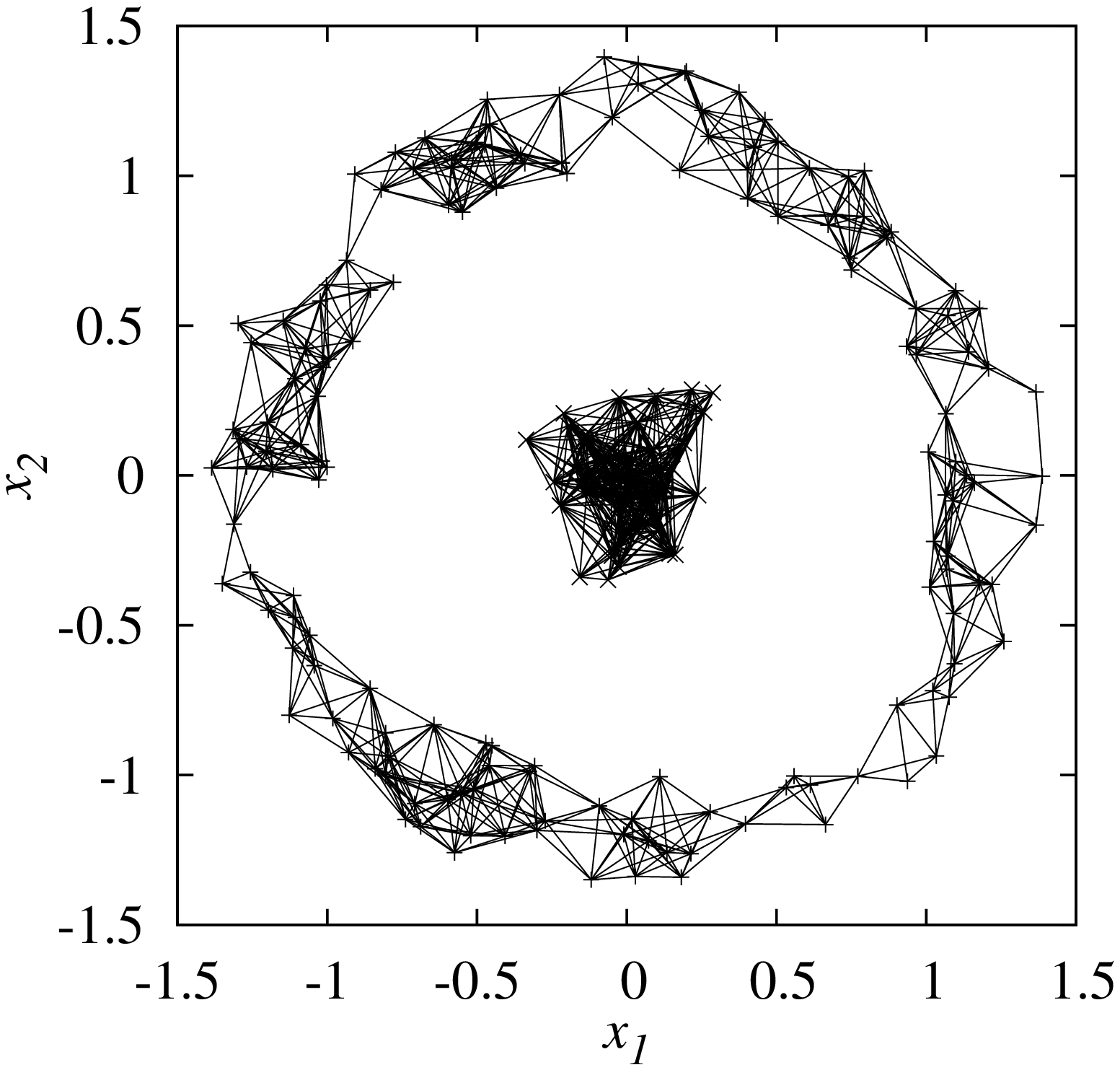}

\includegraphics[width=0.45\textwidth]{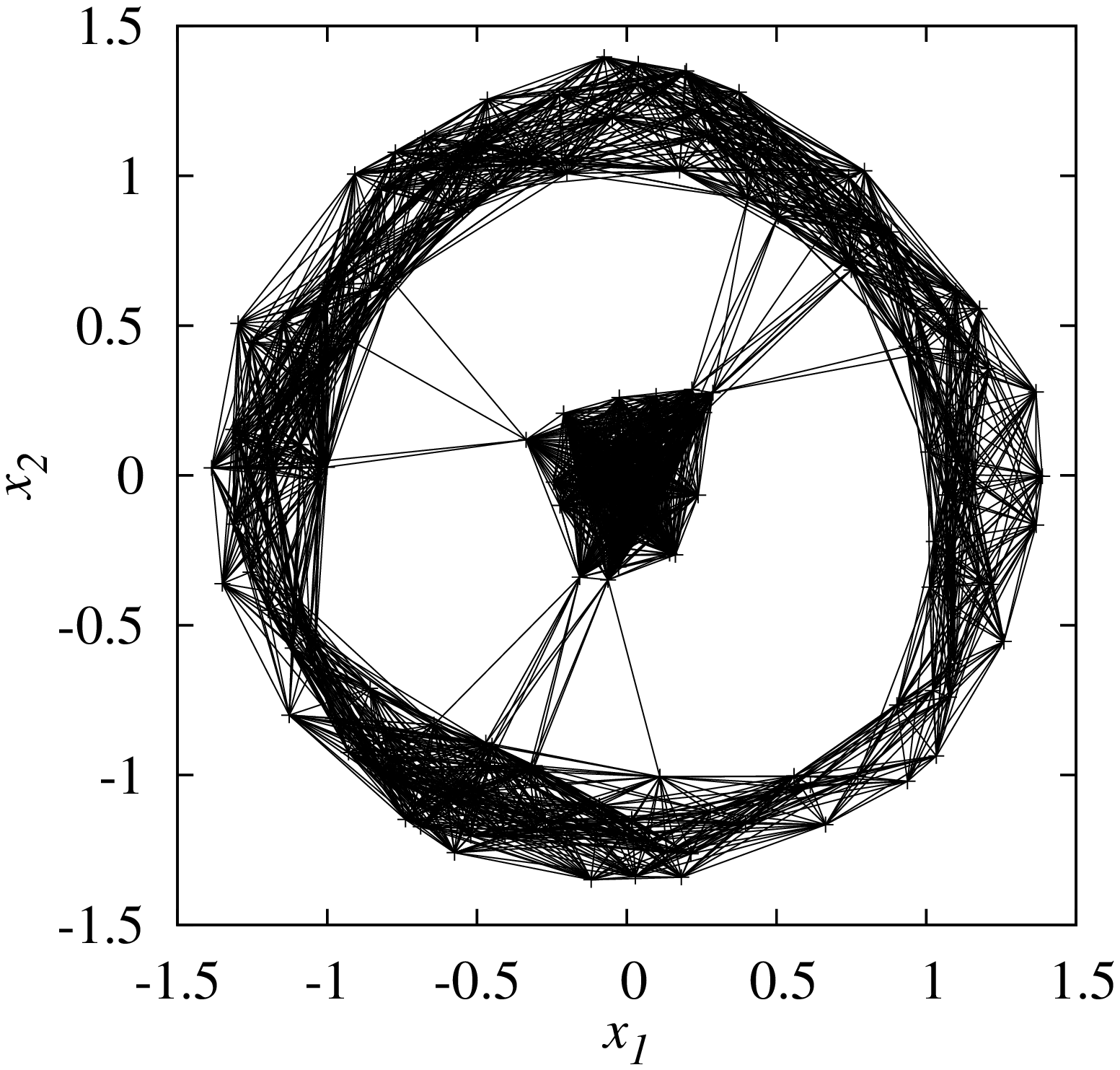}
\includegraphics[width=0.45\textwidth]{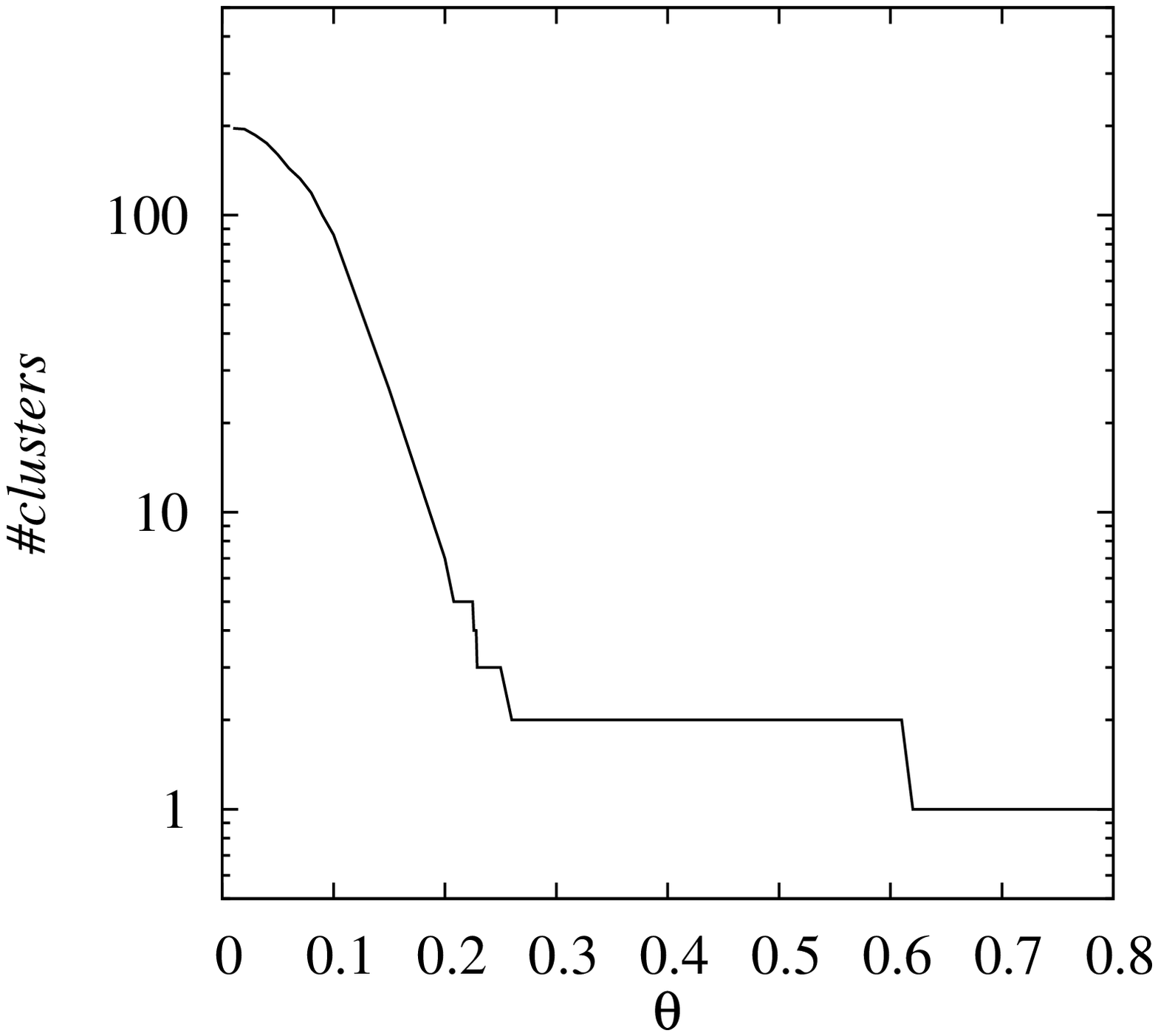}
\end{center}
\caption{
Result of the clustering of sample set B with
the neighbor-based clustering. Upper left: result for threshold $\theta=0.2$.
Upper right: result for threshold $\theta=0.3$.
Lower left: result for threshold $\theta=0.7$.
Lower right: Number of clusters as a function of the threshold $\theta$.
\label{fig:neighbors}}
\end{figure}

\new{Finally, the actual clustering is fairly simple: one just determines
the \emph{connected components}, 
which are the sets of nodes, such that within each
set one can reach from each node all other nodes
of the set via following a finite number edges, called a \emph{path}.
To determine the connected components, the
 function {\tt gs\_components()} is used, which is contained in 
the C source file \texttt{graph\_comps.c}, see also Sec.\ 6.8.4
of Ref.\ \cite{bigPracticalGuide2015}. Now finally,
each connected component corresponds to one cluster!

As example, the neighbor-based clustering 
algorithm is applied to sample set B, where the $k$-means
approach failed. As visible from Fig.\ \ref{fig:neighbors}, the result depends
on the choice of the threshold $\theta$: If the threshold is too small,
too many clusters will be detected, while for a threshold being too large,
just one cluster is found. For intermediate values of the threshold,
the most suitable result of two clusters is found. If the correct threshold
is not known in advance, on can, e.g., study the number of clusters
as a function of the threshold $\theta$. As visible from the
lower right of Fig.\ \ref{fig:neighbors}, the number of clusters
does not change for a large range of thresholds $\theta \in [0.26,0.6]$,
indicating that the most natural number of clusters for sample set B is
two.}

\index{clustering!neighbor-based|)}

\section{Agglomerative Clustering}

\index{clustering!hierarchical|(}
\index{clustering!average-linkage|(}

\new{However, be aware that also neighbor-based clustering might fail. 
Imagine that for sample
set B there is a small ``bridge'' of data points between the two clusters.
In this case, neighbor-based clustering will also not be able to
distinguish the two clusters. In this case, more advanced techniques are 
needed, which are based on the idea that a group of several close-by points 
should influence the outcome of the clustering
 as a group (similar to the $k$ means clustering)
but in terms of distances to other points or groups of points (unlike
$k$-means clustering where only absolute positions are relevant).
 This is the fundamental notion underlying
\emph{hierarchical clustering} methods. These methods are also often able
to detect substructures, like clusters inside clusters etc.
Here, we will focus on an \emph{agglomerative} clustering
approach, namely the \emph{average-linkage} approach.
}

\begin{figure}[!ht]
\begin{center}
\includegraphics[width=0.4\textwidth]{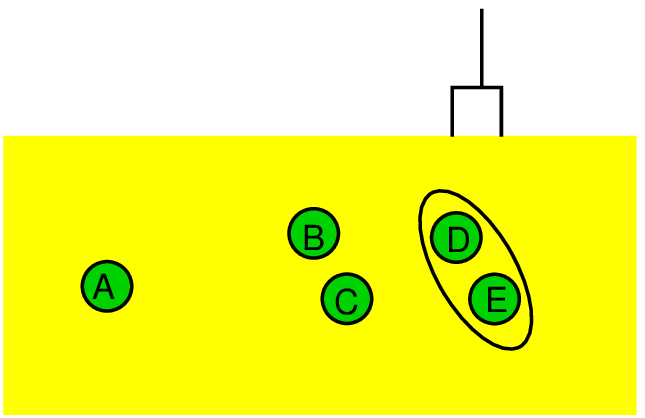}
\includegraphics[width=0.4\textwidth]{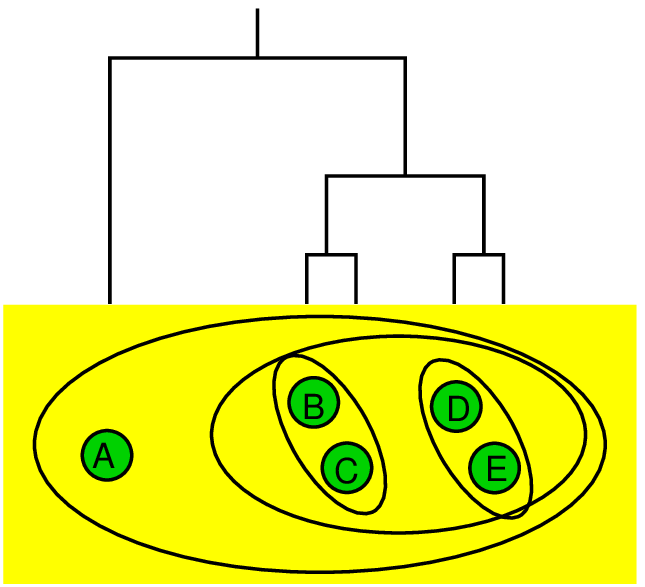}
\end{center}
\caption{
Example for agglomerative clustering: Initially one has a set of $n=5$
data points corresponding to $n$ clusters A, B, C, D, and E (bottom part). 
Iteratively the closest
clusters are merged (illustrated by ellipses). 
For each merger, a branch in a \emph{dendrogram} (tree) is generated (top 
part). Left: Situation after the first two single-point clusters D,E have been
merged into a two-point cluster DE. Right: Final situation, after the merger of
B with C, followed by the merger of BC with DE and finally the merger
of A with BCDE. The 
dendrogram represents the hierarchical cluster structure.
\label{fig:dendo:example}}
\end{figure}

\begin{figure}[!ht]
\begin{center}
\includegraphics[width=0.45\textwidth]{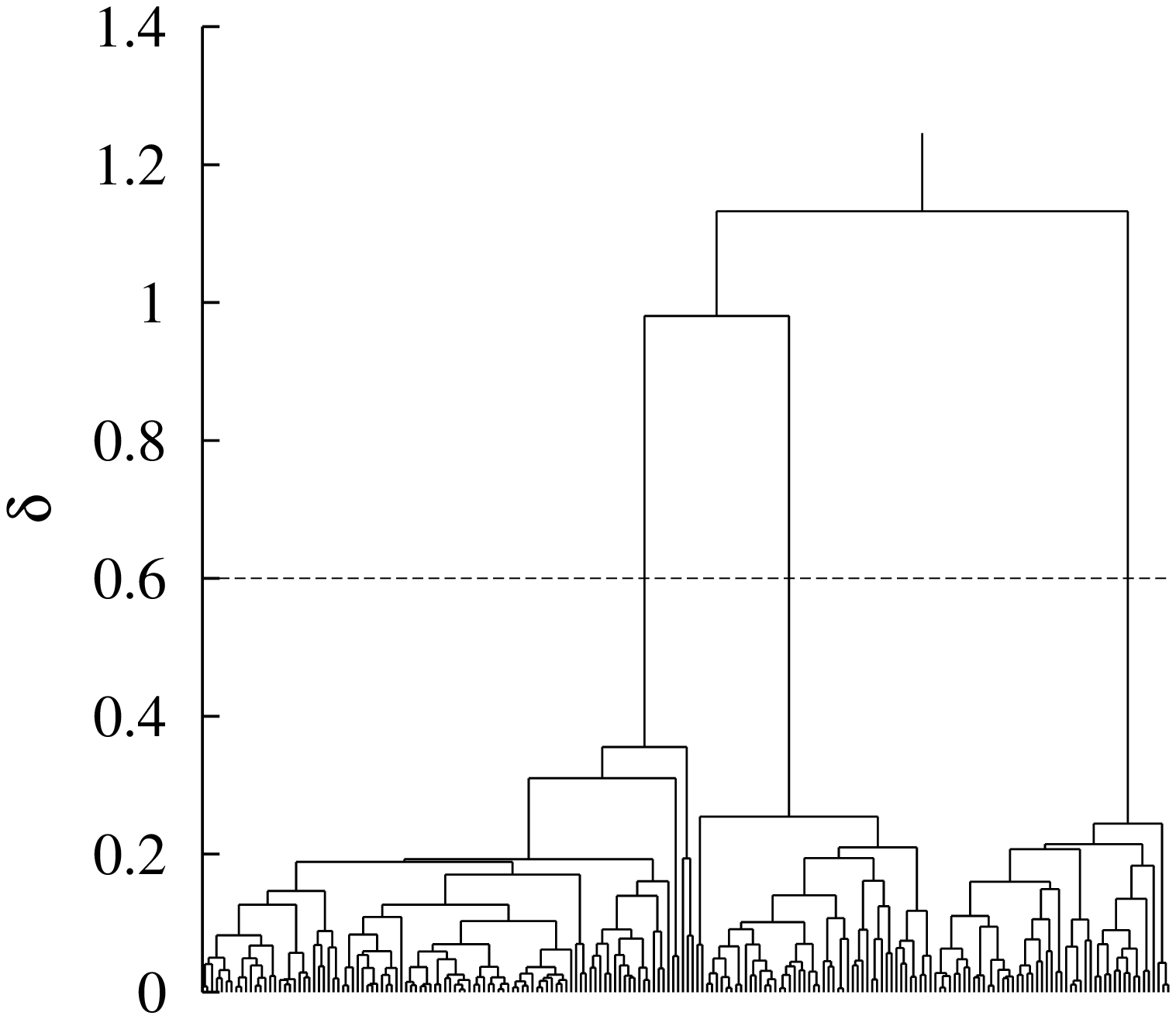}
\includegraphics[width=0.45\textwidth]{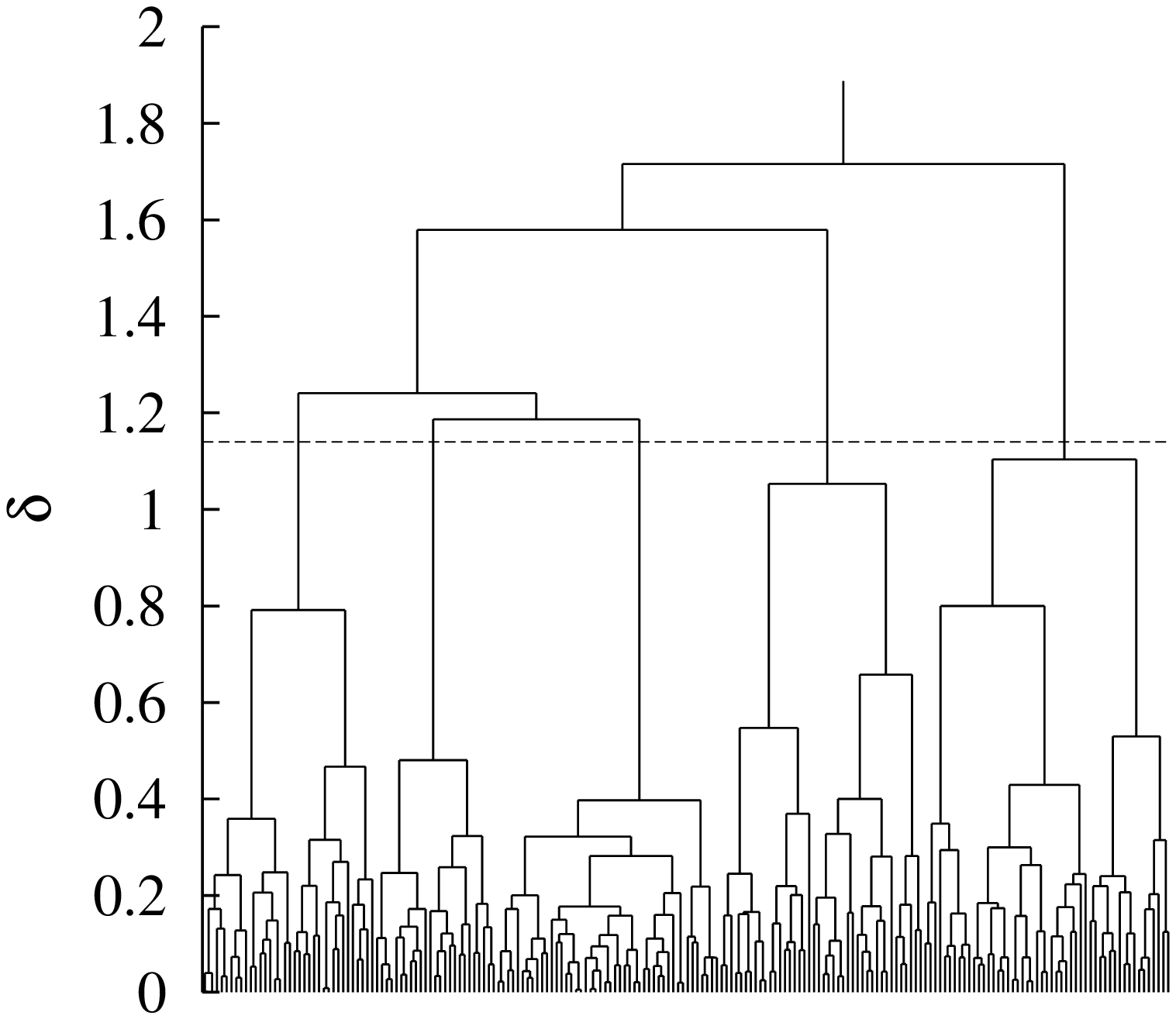}

{\vspace*{2mm}
\includegraphics[width=0.45\textwidth]{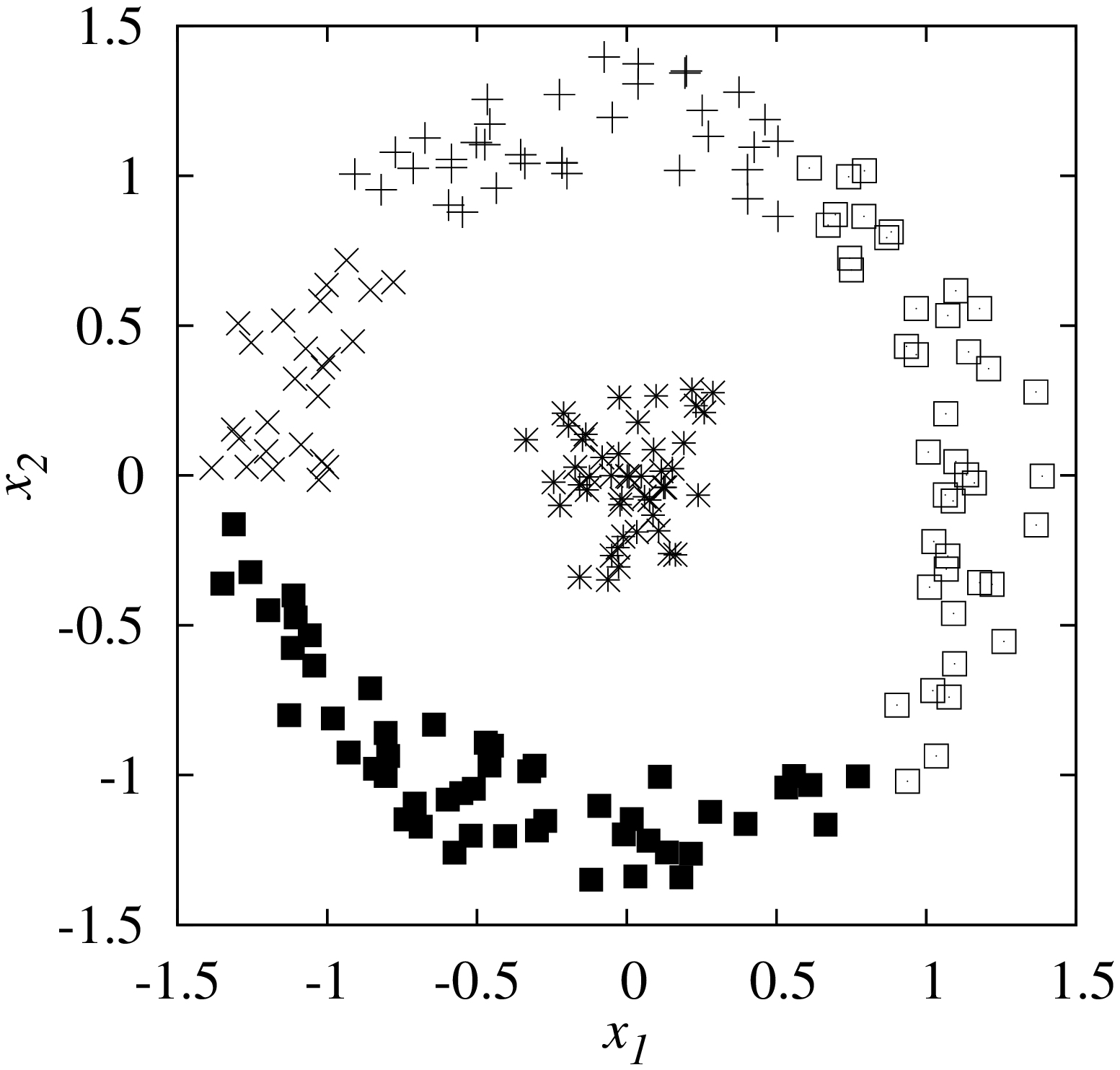}}
{\vspace*{-2mm}
\includegraphics[width=0.45\textwidth]{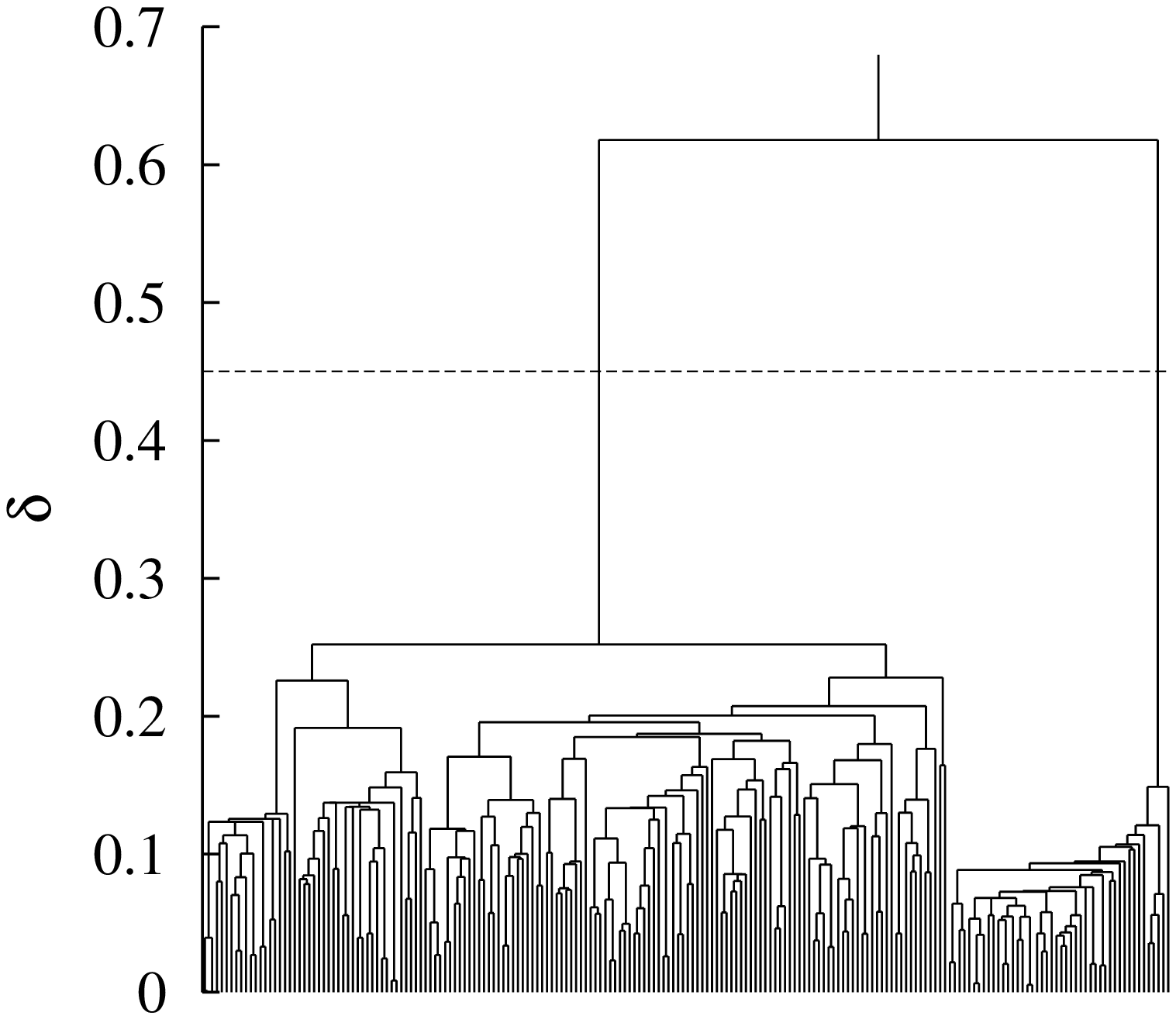}}
\end{center}
\caption{
Results of the clustering of sample set B with
agglomerative clustering. Upper left: dendrogram using average linkage 
clustering for sample set A. Upper right: dendrogram using average linkage 
clustering for sample set B. Lower left: clusters for sample set B
obtained when cutting the dendrogram at height $\delta=1.12$. Lower right:
dendrogram using single-linkage 
clustering for sample set B.
\label{fig:agglomerative:setB}}
\end{figure}

\new{The basic idea of agglomerative clustering 
is that one considers the initial set of $n$ data points
as a set of $n$ clusters $C=\{c_1,\ldots,c_n\}$ with $c_i=\{\myvec{x}^{(i)}\}$. One defines
 cluster distances  $d_c(i,j)$ between pairs of the initial clusters 
$c_i$ and $c_j$ as given by the selected
point-to-point distance function $d(i,j)$, 
like the Euclidean distance Eq.\ (\ref{eq:distance:euclidean})
or any other suitable distance function. Within agglomerative clustering
iteratively the two closest clusters $c_{i_{\min}}$ and $c_{j_{\min}}$,
i.e., where 
$$
i_{\min},j_{\min} = \argmin_{i,j} d_c(i,j)\,,
$$  
are merged into one new single cluster $k= c_{i_{\min}}\cup c_{j_{\min}}$.
Thus, within the first step, two clusters containing a single data point each
will be merged. During the next steps, single-data point clusters or
multiple-data point clusters will be merged. This is illustrated
in Fig.\ \ref{fig:dendo:example}.
During each iteration the number of clusters will be decreased by one,
hence, this process stops after $n-1$ iterations when all data points
are collected in one single cluster. The merging process
can be represented 
by a \index{tree}tree, called \index{dendrogram}\emph{dendrogram}: 
The leaves of the tree are given by the initial data points, i.e.,
the clusters $c_1,\ldots, c_n$. 
Whenever two cluster
are merged, a new (non-leaf) node is created, which has the
two clusters as descendants. Therefore, the root of the tree is the
node which has those two clusters as descendants, which were joined during
the last iteration. Note when drawing the tree, it is convenient
to order the leaves on the $x$-axis according to their appearance
during a  tree traversal, e.g. an \emph{inorder} traversal,
see Sec.\ 6.7 of Ref.\ \cite{bigPracticalGuide2015}.

The most important point is that when creating a cluster $c_k$ through a
merger of $c_{i_{\min}}$ and $c_{j_{\min}}$, 
one has to provide new distances $d_c(k,l)$ of the new cluster $c_k$
to all other clusters $c_l$ with $l\neq i_{\min}$ and $l\neq j_{\min}$.
Different approaches are possible. Here, we use the \emph{average-linking}
clustering, where the distance between two cluster $c_k,c_l$ is the
average distance of the data points in the two clusters:
\begin{equation*}
d_c(k,l) = \frac{1}{|c_k||c_l|} \sum_{i\in c_k, j\in c_l} d(i,j)\,,
\end{equation*}
where $|c_k|$ and $|c_l|$ represent the number of data points in the
clusters $c_k$ and $c_l$, respectively. 
Thus, when cluster $c_k$ is created by merging $c_{i_{\min}}$ and $c_{j_{\min}}$,
the distance of new cluster $c_k$ to all other clusters $c_l$ can be 
conveniently calculated  via
\begin{equation*}
d_c(k,l) = \frac{1}{|c_k|} \left\{|c_{i_{\min}}| d(i_{\min},l)+
|c_{j_{\min}}| d(j_{\min},l)\right\}\,.
\end{equation*}
Many other choices for calculating cluster distances exists, 
 basically they only have to have the property
that the distances between clusters are monotonically increasing when merging.
Common examples are
taking the minimum or the maximum of the point-wise distances between
the nodes of the cluster, specifying \emph{single-linkage}
 and \emph{complete-linkage} 
clustering. Another widely used method is \emph{Ward's} approach, where the
geometric centers of the clusters are also taken into account. 
For details about many clustering algorithms see, e.g., Ref.\ \cite{jain1988}.

Once the clustering procedure is completed
 and the dendrogram calculated, the full
clustering information is contained in the dendrogram, in particular the
hierarchical structure, i.e., if clusters contain sub clusters that in turn
contain sub clusters etc. To obtain a single set of clusters, a common 
approach is to use a threshold $\theta$ such that all inter-cluster
distances are larger than $\theta$ and all intra-cluster distances are 
smaller or equal to
$\theta$. This is similar to the neighbor-based clustering presented before,
only that the intra-cluster distances for agglomerative clustering
 represent joint properties of sub clusters
instead of single pairs of nodes. When drawing the dendrogram, one usually
uses the $\delta=0$ (height) position for the leaves. For all other nodes,
representing mergers of two clusters $i_{\min}, j_{\min}$ , 
one uses a height $\delta\sim d_c(i_{\min}, j_{\min})$, i.e., the distance of the
two clusters  which are merged. Thus, using a threshold $\theta$ corresponds
to drawing a horizontal line at $\delta=\theta$ and cutting off all nodes above
this line, c.f. Fig.\ \ref{fig:agglomerative:setB}.
 The remaining trees located below the line represent the clusters. Often a 
meaningful choice of $\theta$ is to cut the tree at a height value
inside the largest interval where no node has its height in. This correspond
to the iteration where the difference between the distances of the 
last and the current mergers is largest.
}

In the following, we discuss the C implementation of the single-linkage
agglomerative clustering. First, we need a data structure for the nodes
of the dendrogram. Each node stores the ID of the corresponding cluster
and the size of the cluster. If the cluster was merged from two clusters,
the node stores pointers ({\tt left} and {\tt right}) to nodes
corresponding to these clusters
as well as the distance of these two clusters, otherwise the
corresponding entries are NULL (or 0).  For this structure a new
type name {\tt cluster\_node\_t} is introduced:

{\small 
\begin{verbatim}
typedef struct cluster_node
{ 
  int ID;                       /* ID of cluster */
  int size;                 /* number of members */
  double dist;       /* distance of sub clusters */
  struct cluster_node *left;      /* sub cluster */
  struct cluster_node *right;     /* sub cluster */
} cluster_node_t;
\end{verbatim}
}

\new{The function {\tt cluster\_agglomerative()} performs the actual
clustering. It receives a matrix {\tt distance} (GSL type {\tt gsl\_matrix}) 
of point-to-point
distances, as calculated, e.g., by the function {\tt cluster\_distances()}.
The function returns a pointer to the root of the dendrogram, which
represents the clustering. }

\newpage
{\small
\linenumbers[1]
\begin{verbatim}
cluster_node_t *cluster_agglomerative(gsl_matrix *distance)
{
  cluster_node_t *tree;                      /* root of dendrogram */
  cluster_node_t *node;                     /* nodes of dendrogram */
  int num_points;              /* number of points to be clustered */
  int num_clusters;                  /* current number of clusters */
  int next_ID;                               /* ID of next cluster */
  int ID_curr;                            /* ID of current cluster */
  int ID_min1, ID_min2;     /* IDs of clusters having min distance */
  int last_ID;                 /* ID of cluster in last row/column */ 
  int entry_min1, entry_min2;        /* entry  having min distance */
  int c1, c2;                                     /* loop counters */
  int *pos;              /* position of cluster in distance matrix */
  int *cluster;  /* ID of cluster in each row/colum, inv. of 'pos' */
  double delta;                              /* auxiliary distance */
\end{verbatim}
\nolinenumbers}

\begin{figure}[!ht]
\begin{center}
\includegraphics[width=0.55\textwidth]{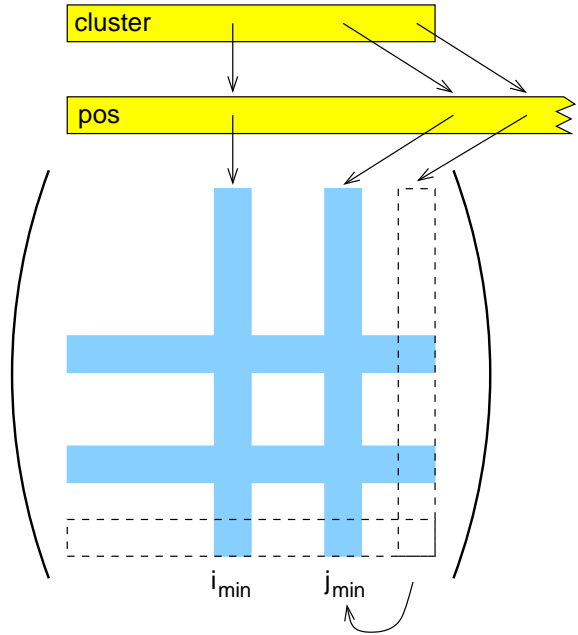}
\end{center}
\caption{When merging clusters with IDs $i_{\min}$ and $j_{\min}$
($i_{\min}<j_{\min}$),
the distances of the new merged cluster are stored in the row
and column where the distances of $i_{\min}$ were stored, while
the entries corresponding to cluster with ID $j_{\min}$ are
swapped with the last row and column. Top part: for each cluster, the current
column and row is stored in the array {\tt pos}, while for
each column and row the current cluster is stored in the array {\tt cluster}. 
\label{fig:cluster:matrix}}
\end{figure}

\noindent
\new{The distances among the data points as well as all 
cluster created during the process 
will be stored in the matrix {\tt distance}. Since there are at most 
$n$ clusters existing at any time, the matrix {\tt distance} is large enough.
When two
clusters are merged, the entries of one cluster will be used
to store the distances of the merged cluster, while the entries
of the other cluster will be disregarded; they will be exchanged
with the distances stored in the last column and row. Thus, after
a merger, the last column and row will not be used any more.
In this way, the matrix {\tt distance} is overwritten.
The current number of used columns and rows, equal to the current
number of clusters is stored in the variable {\tt num\_clusters}.
Note that the cluster IDs are allocated in increasing manner, i.e.,
the IDs 0 to $n-1$ are for the single-data point clusters, the ID
$n$ is for the first cluster created by a merger, the ID $n+1$ for
the second, and so forth.
Since the rows and columns of {\tt distance} contain entries
for all clusters, also for those which are created by mergers, i.e.,
with IDs larger than $n-1$,
two additional arrays are used: The array
{\tt pos} stores for each cluster in which row and column
the corresponding distances are stored currently. 
 Inversely to {\tt pos}, the
array {\tt cluster} stores for each row and column, which
cluster is currently represented there. Thus, we have
always {\tt pos[cluster[i]]==i}
and {\tt cluster[pos[i]]==i}. The data arrangement is illustrated
in Fig.\ \ref{fig:cluster:matrix}. 

In the C code, the number of
data points is determined from the size of the matrix {\tt distance}
(line 16). Next, memory is allocated for the arrays {\tt cluster},
{\tt pos} and {\tt nodes} (line 18--21). The latter two have $2n-1$ entries
since this is the total number of clusters considered during the construction
procedure. The initialization is completed by setting up 
 the entries of {\tt pos} and  {\tt cluster}
and the nodes
for the original data points (lines 23--32):}

{\small
\linenumbers[16]
\begin{verbatim}
  num_points = distance->size1;

  cluster = (int *) malloc(num_points*sizeof(int));
  pos = (int *) malloc( (2*num_points-1)*sizeof(int));
  node = (cluster_node_t *) 
         malloc( (2*num_points-1)*sizeof(cluster_node_t));

\end{verbatim}
\newpage
\begin{verbatim}
  for(c1=0; c1<num_points; c1++)                    /* initialize  */
  {
    pos[c1] = c1;
    cluster[c1] = c1;
    node[c1].left = NULL;
    node[c1].right = NULL;
    node[c1].dist = 0.0;
    node[c1].ID = c1;
    node[c1].size = 1;
  }
\end{verbatim}
\nolinenumbers}

\noindent
\new{
Initially, the number of clusters is equal to the number of data points
$n$ (line 33) and the next available cluster ID will be $n$ (line 34).
The main loop (lines 35--81) will be performed while there
are clusters left for being merged. In the main loop, 
first the smallest current
inter-cluster distance is determined (lines 37--44) and the corresponding
clusters are obtained via the {\tt cluster} array (lines 46,47):}

{\small
\linenumbers[33]
\begin{verbatim}
  num_clusters = num_points;
  next_ID = num_clusters;
  while(num_clusters > 1)        /* until all clusters are merged */
  {
    entry_min1=0; entry_min2=1;  /* search min. off-diag distance */
    for(c1=0; c1<num_clusters; c1++)
      for(c2=c1+1; c2<num_clusters; c2++)
        if(gsl_matrix_get(distance, c1, c2) < 
           gsl_matrix_get(distance, entry_min1, entry_min2))
        {
          entry_min1=c1, entry_min2=c2;
        }

    ID_min1 = cluster[entry_min1];       /* determine cluster IDs */
    ID_min2 = cluster[entry_min2];
\end{verbatim}
\nolinenumbers}

\noindent
\new{Now, a new node can be set up. It contains pointers to its two
sub clusters, its ID, its size which is the sum of the sizes of the two
sub clusters, and the distance of the two sub clusters:}
\newpage

{\small
\linenumbers[48]
\begin{verbatim}
    node[next_ID].left = &(node[ID_min1]);      /* merge clusters */
    node[next_ID].right = &(node[ID_min2]);
    node[next_ID].ID = next_ID;
    node[next_ID].size = node[ID_min1].size + node[ID_min2].size;
    node[next_ID].dist = 
      gsl_matrix_get(distance, entry_min1, entry_min2);
\end{verbatim}
\nolinenumbers}

\noindent
\new{Next, the distances of the remaining clusters to the
new clusters are calculated. These distances are stored in the
entries of the first of the two merged clusters:}

{\small
\linenumbers[54]
\begin{verbatim}
    for(c1=0; c1<num_clusters; c1++)  /* distances to new cluster */
      if(c1 == entry_min1)
        gsl_matrix_set(distance, entry_min1, c1, 0);
      else if(c1 != entry_min2)
      {
        ID_curr = cluster[c1];
        delta = node[ID_min1].size* 
          gsl_matrix_get(distance, entry_min1, c1)+	  
          node[ID_min2].size*
          gsl_matrix_get(distance, entry_min2, c1);
        delta /= node[next_ID].size;
        gsl_matrix_set(distance, entry_min1, c1, delta);
        gsl_matrix_set(distance, c1, entry_min1, delta);
      }
\end{verbatim}
\nolinenumbers}

\noindent
\new{Finally, the current number of clusters is reduced by one (line 68), 
the root of the dendrogram is set if necessary (lines 69 and 70)
the entries of the last current row and column are put
to the row and column where previously the distances of the second cluster
were stored (lines 71--75), the entries of {\tt pos} and {\tt cluster}
for the new cluster are set (lines 77 and 78), and the counter
for the next available cluster ID is increased by one (line 79).
After the main loop has finished, the memory which is associated to those
data structures which are not used any more
is freed (lines 83 and 84):}

\newpage
 
{\small
\linenumbers[68]
\begin{verbatim}
    num_clusters--;
    if(num_clusters == 1)
      tree = &(node[next_ID]);                /* set root of tree */
    last_ID = cluster[num_clusters];/* last cluster -> entry_min2 */
    pos[last_ID] = entry_min2;
    cluster[entry_min2] = last_ID;
    gsl_matrix_swap_rows(distance, num_clusters, entry_min2);
    gsl_matrix_swap_columns(distance, num_clusters, entry_min2);

    cluster[entry_min1] = next_ID;
    pos[next_ID] = entry_min1;
    next_ID++;
	   
  }

  free(pos);                                          /* clean up */
  free(cluster);

  return(tree);
}
\end{verbatim}
\nolinenumbers}

\new{In Fig. \ref{fig:agglomerative:setB} the resulting dendrograms
for sample sets A and B are shown. When cutting the dendrogram for sample
set A at the most obvious height, indeed three clusters emerge. On the
other hand, one has to cut the dendrogram for sample set B at a lower
height to obtain a clustering where the cluster in the middle
is separate from the ``ring'', resulting in five clusters. When considering
a height where four clusters emerge, the ``central'' cluster will be merged
with the cluster to the left indicated by the symbol $\times$,
thus ``ring'' and ``central'' part are not seperated. More
successful is the single-linkage agglomerative approach (not shown here), 
but this is essentially
equivalent to the neighbor-based clustering. The difference
(and improvement) is
that also a dendrogram is obtained which allows to obtain the
most natural threshold and to analyze hierarchical sub structures.

\begin{sloppypar}
Note that the source code file {\tt cluster.c} also contains the function
{\tt cluster\_list\_tree()} which prints for a given dendrogram and
a given threshold $\theta$ the positions of the data points
ordered by the clusters, i.e., between every cluster there will be printed
two empty lines.\footnote{This can be used in {\tt gnuplot} using the
 {\tt index} plot keyword to plot the data points of different clusters
using different symbols.} This function can be easily extended that,
e.g., cluster IDs are assigned to the initial data points.
\end{sloppypar}
}

\index{clustering!average-linkage|)}
\index{clustering!hierarchical|)}
\index{clustering|)}
\index{data clustering|)}

\index{statistical dependence|)}


\begin{thebibliography}{}
\bibitem[Bolobas (1998)]{GR-bollobas98}
Bolobas, B. (1998). {\it Modern Graph Theory}\/,
(Springer, New  York).


\bibitem[Galassi et al.\ (2006)]{gsl}
Galassi M. et al (2006). {\it                                                   
GNU Scientific Library Reference Manual}, (Network Theory Ltd, Bristol),
see also \verb!http://www.gnu.org/software/gsl/!.


\bibitem[Hartmann (2015)]{bigPracticalGuide2015}
Hartmann, A. K. (2015), {\it Big Practical Guide to Computer Simulations},
(World-Scientific, Singapore) 

\bibitem[Jain and Dubes (1988)]{jain1988}
Jain, A. K.  and  Dubes R. C. (1988), 
{\it Algorithms for Clustering Data},
(Prentice-Hall, Englewood Cliffs, USA).

\bibitem[Swamy and Thulasiraman (1991)]{GR-swamy91}
Swamy, M. N. S.  and Thulasiraman, K. (1991).
{\it Graphs, Networks and Algorithms}\/,
(Wiley, New York).

\end{thebibliography}
\end{document}